\documentclass[a4paper,11pt]{article}
\pdfoutput=1 

\usepackage{jcappub} 

\usepackage{multirow}

\title{\boldmath{Prospects for constraining interacting dark energy models from gravitational wave and gamma ray burst joint observation}}

\author[a]{Wan-Ting Hou,}
\author[a,b,1]{Jing-Zhao Qi,}
\author[a,b]{Tao Han,}
\author[a,b]{Jing-Fei Zhang,}
\author[c,d]{Shuo Cao}
\author[a,b,e,f,1]{and Xin Zhang \note{Corresponding author.}}

\affiliation[a]{College of Sciences, Northeastern University, Shenyang 110819, China}
\affiliation[b]{Key Laboratory of Cosmology and Astrophysics (Liaoning Province), Northeastern University, Shenyang 110819, China}
\affiliation[c]{Institute for Frontiers in Astronomy and Astrophysics, Beijing Normal University, Beijing 102206, China}
\affiliation[d]{Department of Astronomy, Beijing Normal University, Beijing 100875, China}
\affiliation[e]{National Frontiers Science Center for Industrial Intelligence and Systems Optimization,
Northeastern University, Shenyang 110819, China}
\affiliation[f]{Key Laboratory of Data Analytics and Optimization for Smart Industry (Ministry of Education), Northeastern University, Shenyang 110819, China}

\emailAdd{houwanting@mail.neu.edu.cn}
\emailAdd{qijingzhao@mail.neu.edu.cn}
\emailAdd{hantao@stumail.neu.edu.cn}
\emailAdd{jfzhang@mail.neu.edu.cn}
\emailAdd{caoshuo@bnu.edu.cn}
\emailAdd{zhangxin@mail.neu.edu.cn}

\abstract{With the measurement of the electromagnetic (EM) counterpart, a gravitational wave (GW) event could be treated as a standard siren. As a novel cosmological probe, GW standard sirens will bring significant implications for cosmology. In this paper, by considering the coincident detections of GW and associated $\gamma$ ray burst (GRB), we find that only about 400 GW bright standard sirens from binary neutron star mergers could be detected in a 10-year observation of the Einstein Telescope and the THESEUS satellite mission. Based on this mock sample, we investigate the implications of GW standard sirens on the interaction between dark energy and dark matter. In our analysis, four viable interacting dark energy (IDE) models, with interaction forms $Q=3\beta H \rho_{\mathrm{de}}$ and $Q=3 \beta H \rho_{\mathrm{c}}$, are considered. Compared with the traditional EM observational data such as CMB, BAO, and SN Ia, the combination of both GW and EM observations could effectively break the degeneracies between different cosmological parameters and provide more stringent cosmological fits. We find that the GW data could play a more important role for determining the interaction in the models with $Q=3 \beta H \rho_{\mathrm{c}}$, compared with the models with $Q=3\beta H \rho_{\mathrm{de}}$.  We also show that constraining IDE models with mock GW data based on different fiducial $H_0$ values yield different results, indicating that accurate determination of $H_0$ is significant for exploring the interaction between dark energy and dark matter. }

\begin{document}
\maketitle
\flushbottom

\section{Introduction}

The last few decades of cosmological study lead us to a standard cosmological scenario, the so-called $\Lambda$ cold dark matter model ($\Lambda$CDM) with six free parameters, which is in remarkable agreement with the bulk of cosmological observations \citep{SupernovaSearchTeam:1998fmf,SupernovaCosmologyProject:1998vns,WMAP:2003elm,SDSS:2003eyi,SDSS:2004wzw,Planck:2015fie,Planck:2018vyg}. This model is featured by dark components including dark matter and dark energy, and dark energy in this model is provided by a cosmological constant $\Lambda$. However, the validity of $\Lambda$CDM has been threatened in recent years on both theoretical and observational grounds. In theory, the cosmological constant $\Lambda$ always suffers from several severe puzzles, such as the ``fine-tuning" and ``cosmic coincidence" problems \citep{Weinberg:1988cp,Sahni:1999gb,Li:2011sd}. In observation, the measurement inconsistencies of some key cosmological parameters, the Hubble constant $H_0$ and cosmic curvature parameter $\Omega_k$ for example, are posing a serious challenge to the standard cosmological model \citep{Planck:2015fie,Riess:2019cxk,Planck:2018vyg,Li:2013dha,Zhao:2017urm,Guo:2018ans,Qi:2020rmm,Qi:2022sxm,Cao:2021zpf,DiValentino:2019qzk,DiValentino:2020hov,Handley:2019tkm}. Therefore, some important cosmological issues need to be reexamined in the context of the serious crisis for the standard cosmology. In fact, all of these problems are relevant to the fundamental natures of dark energy and dark matter. For example, dynamical dark energy with the equation of state (EoS) $w<-1$ can help relieve the $H_0$ tension. The interacting dark energy (IDE) model considering a coupling between dark energy and dark matter not only can alleviate the Hubble tension \citep{Guo:2018ans,Guo:2017hea,Feng:2017usu,Guo:2018gyo,Zhao:2018fjj,Feng:2019mym}, but also can alleviate the ``fine-tuning" and ``cosmic coincidence" problems through the attractor solution \citep{Cai:2004dk,Zhang:2005rg,Zhang:2005rj,Sadjadi:2006qp,Zhang:2007uh}. Additionally, the coupling between dark energy and dark matter is an important theoretical possibility whose confirmation or denial would have enormous implications for fundamental physics. Although the existing observational constraints indicate that the coupling between dark energy and dark matter is weak, they still cannot be ruled out by observations \citep{Costa:2013sva,Nunes:2016dlj,Ferreira:2014jhn,Li:2018ydj,Feng:2019mym,Zhang:2013zyn,2011MNRAS.416.1099C,2011arXiv1105.6274C,2015IJTP...54.1492C,2017JCAP...10..030Z,Guo:2018ans,Guo:2017hea,Feng:2017usu,Guo:2018gyo,Zhao:2018fjj}. Therefore, it is rewarding to test the IDE models by using other complementary cosmological probes including the gravitational wave (GW) observation (especially the inspiraling and merging compact binaries).

The discovery of GW event GW150914 as the first directly detected GW signal marked the arrival of the era of GW astronomy \citep{LIGOScientific:2016aoc,LIGOScientific:2016sjg}. In particular, the later GW170817 event from a binary neutron star (BNS) merger and the successful detections of associate electromagnetic (EM) waves in various bands brought us to a new era of multi-messenger astronomy \citep{LIGOScientific:2017vwq,LIGOScientific:2017zic}. The measurement of GW signal could directly provide the luminosity distance to the source without any additional calibration, and its redshift can be determined by detecting its remaining kilonova emission in the EM band (in the case of BNS), which enables us to establish the luminosity distance-redshift relation \citep{Schutz:1986gp,Zhang:2019ylr,Bian:2021ini}. This is called the standard siren method, which is extremely important for studying cosmology, especially for exploring the constituents of the universe \citep{Zhao:2010sz,Wang:2018lun,Zhang:2018byx,Wang:2019tto,Zhang:2019loq,Zhang:2019ple,Zhao:2019gyk,Qi:2019spg,Qi:2019wwb,Jin:2020hmc,Wang:2021srv,Jin:2021pcv,Qi:2021iic,Jin:2022tdf,Jin:2022qnj,Wu:2022dgy}. However, until now, the only GW bright standard siren event available in cosmology still remains the GW170817. Compared with the current ground-based GW detector network LIGO-Virgo-KAGRA, the upcoming Einstein Telescope (ET) as a third-generation ground-based GW observatory with 10 km-long arms and three detectors has a much wider detection frequency range and a much better detection sensitivity \citep{Maggiore:2019uih}. It is expected to detect abundant standard sirens from the BNS mergers in a 10-year observation \citep{Zhao:2010sz,Cai:2016sby}.

In this paper, we aim to investigate the impacts of the future GW standard siren observations on the constraint of the coupling between dark energy and dark matter. Although some authors have discussed these related issues \citep{Li:2019ajo,Jin:2022tdf,Yang:2019bpr}, we highlight some improvements in this paper as follows.

In previous studies on constraints on cosmology with future GW observations \citep{Li:2019ajo}, the simulated process of GW standard sirens was too rough. It was common to assume that ET could detect 1000 GW standard sirens, which is only a rough estimation. However, determining the redshift for a GW standard siren requires the observation of the EM counterpart. One effective method is to detect a temporally coincident $\gamma$-ray burst (GRB) that can be precisely localized. In this paper, we consider the possibility of simultaneous detection of a joint GW-GRB event. Based on the latest studies about joint GW-GRB observations, we construct a mock catalog of standard sirens from ET by considering the coincidences with a GRB detector. This makes it more reasonable to realistically predict the constraints on the IDE models by the future GW standard siren data.

In the simulated process of GW, the previous studies \citep{Li:2019ajo,Jin:2022tdf,Yang:2019bpr} adopted the $\Lambda$CDM model with the parameters taken from the Planck 2018 results \citep{Planck:2018vyg}, $\Omega_m=0.315$ and $H_0=67.4~\mathrm{km~s^{-1}~Mpc^{-1}}$, as the fiducial cosmology. However, the late-universe observations, the local type Ia supernovae (SN Ia) data calibrated by the distance ladder, reported a high value of the Hubble constant $H_0=73.2\pm1.3~\mathrm{km~s^{-1}~Mpc^{-1}}$ \citep{Riess:2020fzl}, which is in tension with the Planck data. As an independent probe of the late universe, GW data are more likely to infer a $H_0$ value matched with the one inferred by SN Ia data. Therefore, to discuss how robust the constraints on the IDE models from GWs are when the fiducial cosmology is altered, we will simulate the GW data based on two different fiducial $H_0$ values, respectively, and discuss their constraints on the IDE models.

In the previous papers \citep{Li:2019ajo,Jin:2022tdf,Yang:2019bpr}, the interaction terms between dark energy and dark matter are considered proportional to the density of cold dark matter. However, the physical mechanism of the interaction is unclear at present. More discussions of the possibilities would be helpful. In this paper, we will not only consider the interaction terms proportional to the density of cold dark matter but also the interaction terms proportional to the density of dark energy.

Previous studies showed that GW standard sirens, as absolute-distance measurements, are fairly good at constraining the Hubble constant \cite{2020IJMPD..2950105Z} but not good at constraining other cosmological parameters \citep{Qi:2021iic,Zhao:2010sz,Wang:2018lun,Zhang:2019ylr,Wang:2019tto,Zhang:2019loq,Zhang:2019ple,Zhao:2019gyk,Jin:2022tdf,Jin:2022qnj,Jin:2020hmc,Wang:2021srv,Jin:2021pcv,Bian:2021ini,2021ApJ...911..135P,2022RAA....22h5016H}. However, it is found that GW standard sirens are highly complementary with some conventional cosmological probes, including the cosmic microwave background (CMB), baryon acoustic oscillations (BAO), and SN Ia in constraining dark energy models. Such data combinations could effectively break the parameter degeneracies and give tighter constraints on cosmological parameters. Therefore, in this paper, we also employ the mainstream cosmological probes combined with the simulated GW standard siren data from ET to thoroughly investigate the GW's role in studying the IDE models. The traditional observational data sets used in this paper include CMB, BAO, and SN Ia. Compared to the previous works, we will employ the latest SN Ia ``Pantheon+" compilation \citep{Brout:2022vxf}, containing 1701 light curves of 1550 unique objects, instead of the ``Pantheon" data containing 1048 data points. There are many improvements, including the sample size, the treatments of systematic uncertainties in redshift, peculiar velocities, photometric calibration, and intrinsic scatter model of SN Ia, which greatly enhance the constraining capability of the Pantheon+ compilation compared with the original Pantheon compilation \citep{Brout:2022vxf}. Therefore, our work in this paper includes the latest constraints on the IDE model from traditional observational data. For the CMB measurements, we use the Planck distance priors $\left(R, \ell_{A}, \Omega_{b} h^{2}\right)$ obtained from the Planck 2018 TT,TE,EE+lowE data \citep{Planck:2018vyg,Chen:2018dbv}. For the BAO data, we consider several measurements from 6dFGS \citep{Beutler:2011hx}, SDSSMGS \citep{Ross:2014qpa}, and BOSS DR12 \citep{BOSS:2016wmc}.

\section{Interacting dark energy models}

In a flat universe described by the Friedmann-Lema\^{\i}tre-Robertson-Walker metric, the dimensionless Hubble parameter $E(z)=H(z)/H_0$ could be written as
\begin{equation}
E^{2}=\Omega_{\mathrm{de} 0} \frac{\rho_{\mathrm{de}}}{\rho_{\mathrm{de} 0}}+\Omega_{\mathrm{c} 0} \frac{\rho_{\mathrm{c}}}{\rho_{\mathrm{c} 0}}+\Omega_{\mathrm{b} 0} \frac{\rho_{\mathrm{b}}}{\rho_{\mathrm{b} 0}}+\Omega_{\mathrm{r} 0} \frac{\rho_{\mathrm{r}}}{\rho_{\mathrm{r} 0}},
\end{equation}
where $\Omega_{\mathrm{de} 0}$, $\Omega_{\mathrm{c} 0}$, $\Omega_{\mathrm{b} 0}$ and $\Omega_{\mathrm{r} 0}$ are current energy density fractions of dark energy, cold dark matter (CDM), baryon and radiation, respectively. Here, we have $\rho_{\mathrm{b}}=\rho_{\mathrm{b} 0}(1+z)^{3}$ and $\rho_{\mathrm{r}}=\rho_{\mathrm{r} 0}(1+z)^{4}$, where $z$ denotes the redshift. Considering the interaction between dark energy and CDM, we have the energy continuity equations 
\begin{equation}
\begin{aligned}
&(1+z) \frac{\mathrm{d} \rho_{\mathrm{de}}}{\mathrm{d} z}-3(1+w) \rho_{\mathrm{de}}=\frac{Q}{H}, \\
&(1+z) \frac{\mathrm{d} \rho_{\mathrm{c}}}{\mathrm{d} z}-3 \rho_{\mathrm{c}}=-\frac{Q}{H},
\end{aligned}
\end{equation}
where $Q$ denotes a phenomenological interaction term, and $w$ is the EoS of dark energy. The form of $Q$ is an open question. However, it is usually assumed to be proportional to the density of dark sectors. In this paper, we consider two forms of the interaction term, i.e., $Q_1=3\beta H \rho_{\mathrm{de}}$ and $Q_{2}=3 \beta H \rho_{\mathrm{c}}$, where $\beta$ is a dimensionless coupling parameter describing the strength of the interaction between dark energy and dark matter. $\beta>0$ means that dark matter will be converted into dark energy, and vice versa for $\beta<0$. If $\beta=0$, it indicates no interaction between the two sectors.

For the EoS of dark energy $w$, we consider two cases, i.e., the case of $w=-1$, denoted as I$\Lambda$CDM, and the case of $w$ being a constant, denoted as I$w$CDM. Thus, we will have four IDE models, I$\Lambda$CDM1, I$\Lambda$CDM2, I$w$CDM1, and I$w$CDM2. For the I$w$CDM1 model (with $Q_1=3\beta H \rho_{\mathrm{de}}$), the dimensionless Hubble parameter satisfies \citep{Wang:2014fqa,2011arXiv1105.6274C,Xia:2016vnp}
\begin{equation}
\begin{aligned}
E^{2}(z)=& \Omega_{\mathrm{de} 0}\left(\frac{\beta}{w+\beta}(1+z)^{3}+\frac{w}{w+\beta}(1+z)^{3(1+w+\beta)}\right) \\
&+\Omega_{\mathrm{m} 0}(1+z)^{3}+\Omega_{\mathrm{r} 0}(1+z)^{4}.
\end{aligned}
\end{equation}
For the I$w$CDM2 model (with $Q_{2}=3 \beta H \rho_{\mathrm{c}}$), we have \citep{Wang:2014fqa,2011arXiv1105.6274C,Xia:2016vnp}
\begin{equation}
\begin{aligned}
E^{2}(z)=& \Omega_{\mathrm{de} 0}(1+z)^{3(1+w)}+\Omega_{\mathrm{b} 0}(1+z)^{3}+\Omega_{\mathrm{r} 0}(1+z)^{4} \\
&+\Omega_{\mathrm{c} 0}\left(\frac{\beta}{w+\beta}(1+z)^{3(1+w)}+\frac{w}{w+\beta}(1+z)^{3(1-\beta)}\right).
\end{aligned}
\end{equation}
Setting $w=-1$ in the above two expressions, the corresponding expressions for I$\Lambda$CDM1 and I$\Lambda$CDM2 could be obtained.

\section{Gravitational wave simulation}

In this paper, we wish to realistically forecast the constraints on the IDE models from GW standard sirens detected by ET. To construct a mock catalogue of standard sirens, not only the detectable mergers of BNS should be considered, but also more importantly, whether the associated EM counterparts are able to be detected should be considered. Therefore, in this work, we consider Transient High-Energy Sky and Early Universe Surveyor (THESEUS) mission \citep{THESEUS:2017qvx,THESEUS:2017wvz,Stratta:2018ldl},  a space telescope proposed to study GRB and X-rays, to predict the associated GRBs with GWs.

Based on the star formation rate \citep{Vitale:2018yhm,Belgacem:2019tbw, Yang:2021qge}, the merger rate of BNS density per unit redshift is
\begin{equation}
R_z(z)=\frac{R_m(z)}{(1+z)}\frac{dV(z)}{dz},
\end{equation}
where $dV/dz$ is the comoving volume element. $R_m(z)$ represents the rate per volume in the source frame, which is related to the time delay between the formation of the BNS progenitors and their mergers as
\begin{equation}
R_m(z)=\int_z^{\infty}\frac{dt_f}{dz_f}R_f(z_f)P(t_d)dz_f,
\end{equation}
where $R_f$ is the formation rate of massive binaries and $P(t_d)$ is the distribution of the time delay $t_d$.  A BNS system that merges at the look-back time $t$ is formed at the look-back time $t_f$, and the time delay $t_d$ is $t_f-t$. Here, the formation rate of massive binaries $R_f$ is assumed to be proportional to the cosmic star formation rate, for which we adopt the Madau-Dickinson model \citep{Madau:2014bja},
\begin{equation}
\psi_{\rm{MD}}=\psi_0\frac{(1+z)^{\alpha^{\prime}}}{1+[(1+z)/C]^{\beta^{\prime}}},
\end{equation}
with parameters $\alpha^{\prime}=2.7$, $\beta^{\prime}=5.6$, $C=2.9$ and $\psi_0=0.015~ \rm{M_{\odot}~Mpc^{-3}yr^{-1}}$. The proportionality coefficient is the normalization factor that ensures today the merger rate is $R_m(z=0)=R_0$,  for which we use $R_0=920~\rm{Gpc^{-3}yr^{-1}}$ estimated from the O1 LIGO and O2 Ligo/Virgo observation run \citep{LIGOScientific:2018mvr}. For the time delay distribution $P(t_d)$, we follow and adopt the exponential form \citep{Vitale:2018yhm}, 
\begin{equation}
P(t_d)=\frac{1}{\tau}\exp(-t_d/\tau),
\end{equation}
with time scale parameter $\tau=0.1~\rm{Gyr}$. Thus, we can obtain the BNS merger rate density per unit redshift.

\begin{figure*}
\centering
\includegraphics[scale=0.4]{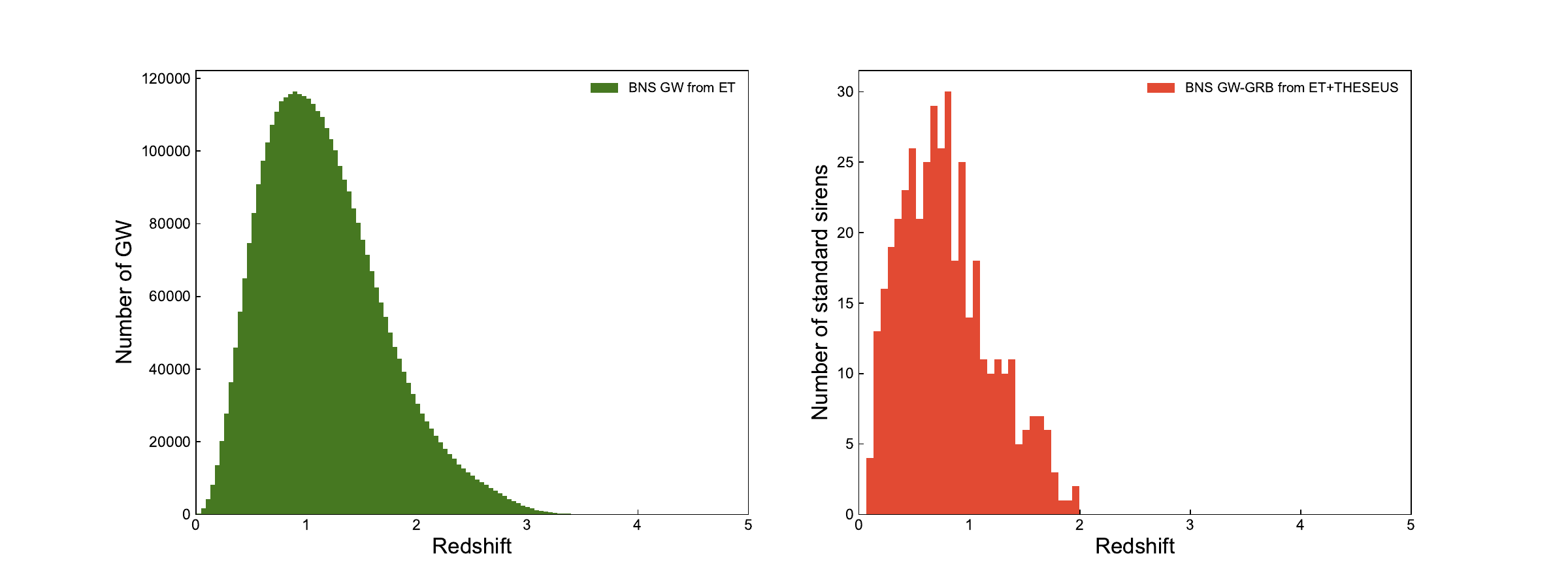}
\caption{The redshift distribution of the mock catalogue for a 10-year detection of BNS GW (left) from ET and GW-GRB standard sirens (right) from ET+THESEUS. \label{gwgrb}}
\end{figure*}

To obtain the detectable number and distribution of GW events, we need to calculate the signal-to-noise ratio (SNR) for each GW event and select the event whose SNR is larger than the sensitivity threshold of the GW detector. For the calculation of the SNR for a GW event, we refer to the detailed description in Refs. \citep{Zhao:2010sz,Cai:2016sby}, and we do not repeat it here. For the threshold of ET, we take it to be $\rho_{\rm{threshold}}=8$. We assume a running period of 10 years and a duty cycle of 80\%. In the left panel of figure \ref{gwgrb}, we show the redshift distribution of GW events from a realization of the mock catalogue for the ET in a 10-year observation.

For the BNSs used as standard sirens, the redshift information of sources is necessary, which could usually be inferred by the observations of electromagnetic counterparts. To estimate the number and distribution of coincidences between GW events and EM counterparts, the available network of GRB satellites and telescopes when the GW detector is triggered is very crucial. In the following, we combine the THESEUS mission to calculate the detection rate of GW-GRB standard sirens.

For a GRB detected in coincidence with a GW signal, the peak flux should be larger than the flux limit of the satellite. Based on the analysis of GRB170817A \citep{Howell:2018nhu}, we adopt the Gaussian structured jet profile model,
\begin{equation}
L\left(\theta_{\mathrm{V}}\right)=L_{\mathrm{c}} \exp \left(-\frac{\theta_{\mathrm{V}}^2}{2 \theta_{\mathrm{c}}^2}\right),
\end{equation}
where $L\left(\theta_{\mathrm{V}}\right)$ is the luminosity per unit solid angle, $\theta_{\mathrm{V}}$ is the viewing angle, $L_{\mathrm{c}}$ and $\theta_{\mathrm{c}}$ are structure parameters defined the angular profile. The structured jet parameter is given by $\theta_{\mathrm{c}}=4.7^{\circ}$. $L_{\mathrm{c}}=L_{\mathrm{p}} / 4 \pi$ {erg} {s}$^{-1}$ {sr}$^{-1}$, where $L_{\rm{p}}$ is the peak luminosity of each burst. Next, to determine if a GRB could be detected, we need to convert the flux threshold of the GRB satellite $P_{\rm{T}}$ to the luminosity, which could be implemented by 
\begin{equation}
L=4 \pi d_L^2(z) k(z) b /(1+z) P_T.
\end{equation}
$b$ is  an energy normalization to account for the missing fraction of the $\gamma$-ray energy seen in the detector band \citep{Wanderman:2014eza,Howell:2018nhu}, and its expression is
\begin{equation}
b=\frac{\int_{10000}^1 E N(E) d E}{\int_{E 1}^{E 2} N(E) d E},
\end{equation}
where $N(E)$ is the observed GRB photon spectrum in units of $\rm{ph~s^{-1}~keV^{-1}~cm^{-2}}$, and $[E1,E2]$ is the detector's energy window. $k(z)$ represents a k-correction \citep{Wanderman:2014eza,Howell:2018nhu},
\begin{equation}
k(z)=\frac{\int_{E 1}^{E 2} N(E) d E}{\int_{E 1(1+z)}^{E 2(1+z)} N(E) d E}.
\end{equation}
For the observed GRB photon spectrum, we model it by the Band function, which is a function of spectral indices ($\alpha_B$, $\beta_B$) and break energy $E_b$, expressed as \citep{Band:2002te}
\begin{equation}
N(E)= \begin{cases}N_0\left(\frac{E}{100 \mathrm{keV}}\right)^{\alpha_B} \exp \left(-\frac{E}{E_0}\right), & E \leq E_b \\ N_0\left(\frac{E_b}{100 \mathrm{keV}}\right)^{\alpha_B-\beta_B} \exp \left(\beta_B-\alpha_B\right)\left(\frac{E}{100 \mathrm{keV}}\right)^{\beta_B}, & E>E_b\end{cases},
\end{equation}
where $E_b=\left(\alpha_B-\beta_B\right) E_0$ and $E_p=\left(\alpha_B+2\right) E_0$.  We take $\alpha_B=-0.5$, $\beta_B=-2.25$ and a peak energy $E_p=800~\rm{keV}$.

For the distribution of the short GRBs, we assume a standard broken power law model,
\begin{equation}
\Phi\left(L\right) \propto \begin{cases}\left(L/ L_*\right)^{\alpha_L}, & L<L_* \\ \left(L / L_*\right)^{\beta_L}, & L \geq L_*\end{cases},
\end{equation}
with the characteristic parameter separating the two regimes $L_*=2\times10^{52}~\rm{erg~sec^{-1}}$, and two slopes parameters $\alpha_L=-1.95$ and $\beta_L=-3$ \citep{Wanderman:2014eza}.  For the THESEUS mission \citep{THESEUS:2017wvz}, it is recorded if the value of the observed flux is larger than the flux threshold $P_{\rm{T}}=0.2~\rm{ph~sec^{-1}~cm^{-2}}$ in the $50-300$ keV band. According to the THESEUS paper \citep{THESEUS:2017wvz}, we take a sky coverage fraction of 0.5 and a duty cycle of 80\%. Then we can calculate the probability of the GRB detection for every GW event according to the probability distribution $\Phi(L)dL$. Finally, we find that only about 400 standard sirens could be detected for the ET+THESEUS network in 10 years, whose redshift distribution is shown in the right panel of figure \ref{gwgrb}.

In order to discuss the influence of GW simulation based on different fiducial $H_0$ values on the final results, two sets of GW data will be simulated here. We adopt two fiducial cosmological models, both of which are the $\Lambda$CDM models, but with different parameter values. One model uses the parameter values from Planck 2018 results \citep{Planck:2018vyg}, namely $\Omega_{\rm{m}}=0.315$ and $H_0=67.4 ~\mathrm{km~s^{-1}~Mpc^{-1}}$, while the other model use $\Omega_{\rm{m}}=0.315$ from Planck 2018, but $H_0=73.2~\mathrm{km~s^{-1}~Mpc^{-1}}$ from SH0ES Collaboration \citep{Riess:2020fzl}. According to the error strategy, we consider the instrumental error $\sigma_{D_L}^{\text {inst }}$ and an additional error $\sigma_{D_L}^{\text {lens }}$ caused by the weak lensing to the total uncertainty of the luminosity distance as
\begin{equation}
\sigma_{D_L}=\sqrt{\left(\sigma_{D_L}^{\text {inst }}\right)^2+\left(\sigma_{D_L}^{\text {lens }}\right)^2}.
\end{equation}
The instrumental error of the luminosity distance is dependent on the signal-to-noise ratio (SNR) $\rho$ of a GW signal as $\sigma_{D_L}^{\text{inst}} \simeq D_L / \rho$ \citep{Zhao:2010sz}. For the ET, to confirm a detected GW signal is using the criteria that the combined SNR is larger than 8 \citep{Zhao:2010sz,Cai:2016sby}. To consider the correlation between the inclination angle $\iota$ and the luminosity distance, we add a factor 2 in front of the error \citep{Cai:2016sby}, $\sigma_{D_L}^{\text{inst}} \simeq 2D_L / \rho$. The lensing uncertainty is modelled as $\sigma_{D_L}^{\text {lens }}=0.05 z D_L$ \citep{Cai:2016sby}.

\section{Results and discussions}

\begin{figure*}
\centering
\includegraphics[scale=0.45]{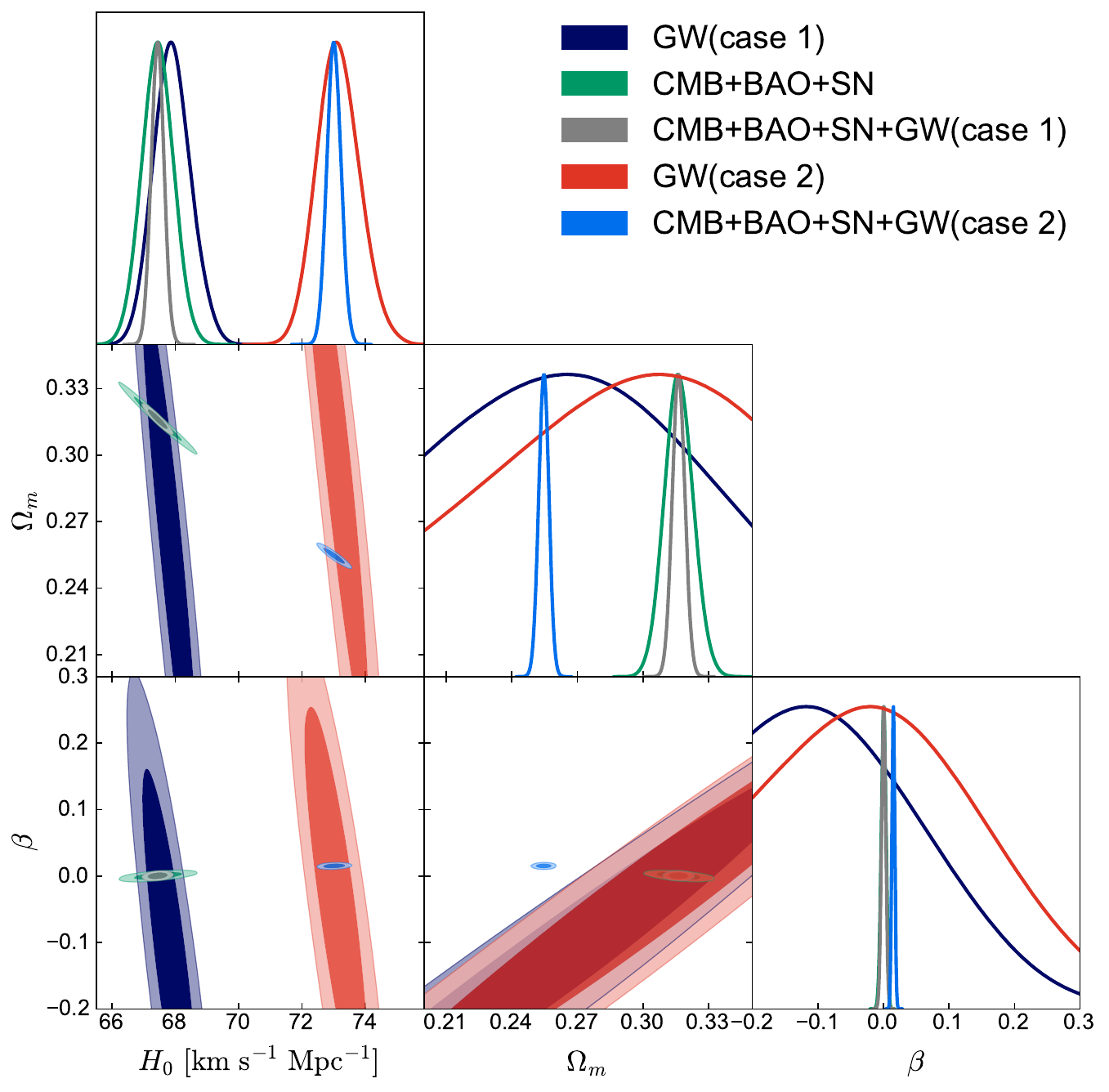}
\caption{The constraints (68.3\% and 95.4\% confidence level) on the I$\Lambda$CDM1 model with $Q=3\beta H\rho_{\rm{de}}$ from GW, CMB+BAO+SN, and CMB+BAO+SN+GW. Case 1 denotes the simulated GW data with $H_0=67.4~\mathrm{km~s^{-1}~Mpc^{-1}}$, and case 2 represents the one with $H_0=73.2~\mathrm{km~s^{-1}~Mpc^{-1}}$, respectively. \label{lcdm1}}
\end{figure*} 

\begin{figure*}
\centering
\includegraphics[scale=0.45]{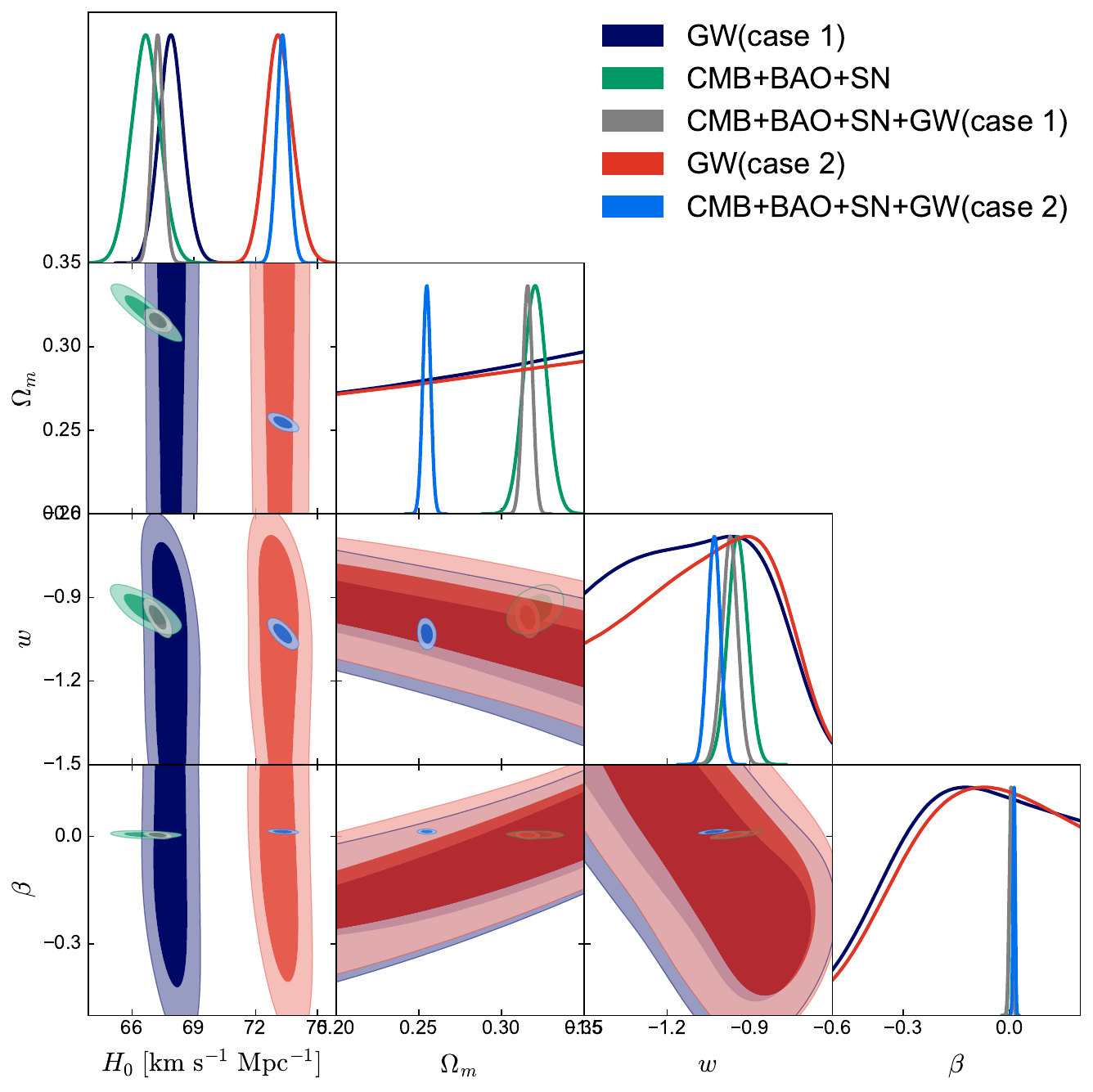}
\caption{Same as figure \ref{lcdm1} but for the I$w$CDM1 model with $Q=3\beta H\rho_{\rm{de}}$. \label{wcdm1}}
\end{figure*}   

\begin{figure*}
\centering
\includegraphics[scale=0.45]{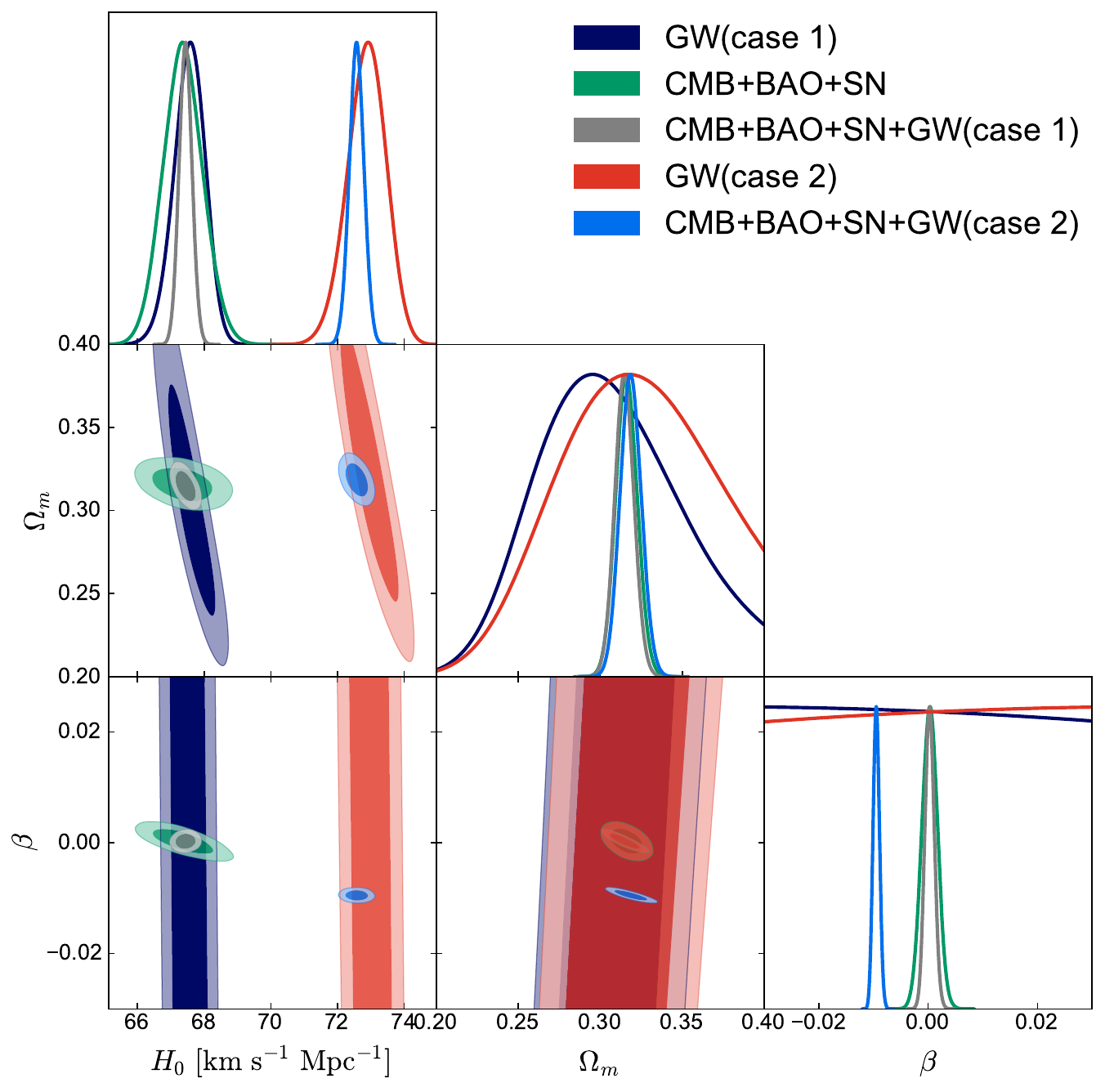}
\caption{Same as figure \ref{lcdm1} but for the I$\Lambda$CDM2 model with $Q=3\beta H\rho_{\rm{c}}$. \label{lcdm2}}
\end{figure*}   

\begin{figure*}
\centering
\includegraphics[scale=0.45]{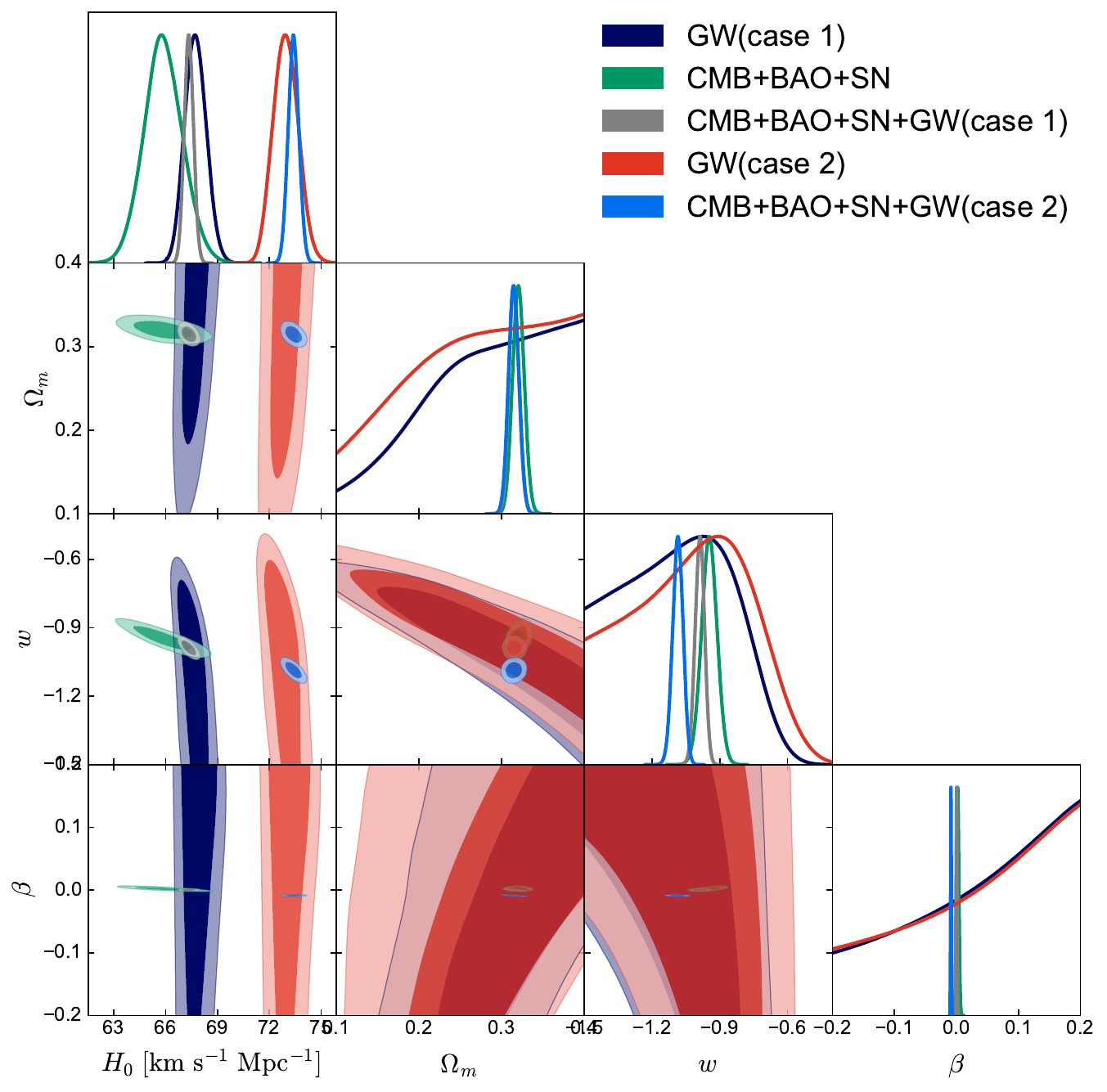}
\caption{Same as figure \ref{lcdm1} but for the I$w$CDM2 model with $Q=3\beta H\rho_{\rm{c}}$. \label{wcdm2}}
\end{figure*}

\begin{table}[tbp]
\centering
\renewcommand{\arraystretch}{1.5}
\resizebox{\textwidth}{!}{
\begin{tabular}{lcccccc}
\hline
\textbf{Model} & \textbf{Parameter} &\textbf{CMB+BAO+SN}  &\textbf{GW(case 1)}&\textbf{CMB+BAO+SN+GW(case 1)}&\textbf{GW(case 2)}&\textbf{CMB+BAO+SN+GW(case 2)} \\  \hline
\multirow{3}{*}{\shortstack{I$\Lambda$CDM1 \\ ($Q_1=3\beta H\rho_{\rm{de}}$)}}
    &$H_{0}$            &$67.44\pm{0.50}$    &$67.90^{+0.56}_{-0.62}$       &$67.45\pm{0.20}$  &$73.15^{+0.63}_{-0.72}$       &$73.01\pm{0.22}$\\
    &$\Omega_{\rm m}$   &$0.3164\pm{0.0066}$  &$0.251^{+0.088}_{-0.071}$     &$0.3163\pm{0.0029}$ &$0.290^{+0.093}_{-0.072}$     &$0.2546\pm{0.0023}$ \\
    &$\beta$            &$0.0003\pm{0.0036}$    &$-0.15^{+0.21}_{-0.18}$    &$-0.0003\pm{0.0032}$&$-0.06^{+0.22}_{-0.18}$ &$0.0152\pm{0.0022}$\\
    \hline
\multirow{4}{*}{\shortstack{I$w$CDM1\\ ($Q_1=3\beta H\rho_{\rm{de}}$)}}
    &$H_{0}$ &$66.66\pm{0.70}$ &$67.90\pm 0.60$&$67.25\pm{0.28}$ &$72.92\pm 0.70$ &$73.32\pm{0.30}$ \\
    &$\Omega_{\rm m}$ &$0.3203\pm 0.0071$ &$ >0.266$ &$0.3157\pm{0.0029}$ &$ 0.37^{+0.19}_{-0.12}$      &$0.2546\pm{0.0023}$\\
    &$w$ &$-0.944\pm 0.036$ &$-1.23^{+0.43}_{-0.28}$ &$-0.972\pm{0.029}$ &$-1.14^{+0.45}_{-0.22}$ &$-1.030\pm{0.023}$\\
    &$\beta$ &$0.0012\pm 0.0045$ &$0.08^{+0.30}_{-0.39}$ &$0.0008^{+0.0045}_{-0.0041}$&$0.11^{+0.39}_{-0.23}$ &$0.0131\pm 0.0029$\\
    \hline
\multirow{3}{*}{\shortstack{I$\Lambda$CDM2\\ ($Q_2=3\beta H\rho_{\rm{c}}$)}}
    &$H_{0}$ &$67.37\pm 0.60$ &$67.55^{+0.50}_{-0.43}$ &$67.44\pm{0.19}$ &$72.87^{+0.61}_{-0.54}$ &$72.57\pm{0.22}$\\
    &$\Omega_{\rm m}$ &$0.3160\pm 0.0065$ &$0.312^{+0.035}_{-0.057}$ &$0.3148\pm{0.0059}$&$0.335^{+0.044}_{-0.068}$ &$0.3186\pm{0.0065}$ \\
    &$\beta$ &$0.0003\pm 0.0015$ &$0.01^{+0.24}_{-0.32}$ &$0.0003\pm{0.0009}$ &$0.08^{+0.24}_{-0.30}$ &$0.0095\pm{0.0006}$\\
    \hline
    \multirow{4}{*}{\shortstack{I$w$CDM2\\ ($Q_2=3\beta H\rho_{\rm{c}}$)}}
    &$H_{0}$   &$65.9\pm{1.1}$ &$67.74\pm{0.62}$ &$67.34\pm{0.27}$ &$72.96\pm{0.72}$ &$73.41\pm{0.31}$\\
    &$\Omega_{\rm m}$ &$0.3202\pm{0.0071}$&$0.388^{+0.18}_{-0.091}$&$0.3148\pm{0.0060}$&$>0.291$&$0.3146\pm{0.0064}$ \\
    &$w$  &$-0.946\pm{0.035}$ &$-1.26^{+0.47}_{-0.28}$ &$-0.988\pm{0.021}$&$-1.16^{+0.47}_{-0.28}$&$-1.086\pm{0.023}$\\
    &$\beta$ &$0.0020\pm{0.0019}$&$0.12^{+0.36}_{-0.20}$ &$0.0002\pm{0.0009}$&$>0.0414$&$-0.0089\pm{0.0006}$\\
\hline
\end{tabular}}
\caption{The constraints on cosmological parameters of the I$\Lambda$CDM1, I$w$CDM1, I$\Lambda$CDM2 and I$w$CDM2 models from GW, CMB+BAO+SN and CMB+BAO+SN+GW. Case 1 denotes the simulated GW data with $H_0=67.4~\mathrm{km~s^{-1}~Mpc^{-1}}$, and case 2 represents the one with $H_0=73.2~\mathrm{km~s^{-1}~Mpc^{-1}}$, respectively. Here $H_{0}$ is in units of $\rm km\ s^{-1}\ Mpc^{-1}$.\label{results_table}}
\end{table}

The main results are shown in figures \ref{lcdm1}--\ref{wcdm2} and table \ref{results_table}. Figure \ref{lcdm1} shows the constraints on the I$\Lambda$CDM1 model with $Q=3\beta H\rho_{\rm{de}}$ from GW, CMB+BAO+SN and CMB+BAO+SN+GW. It is important to emphasize that the central values of the cosmological parameters obtained from the simulated GW data are not meaningful in themselves, and only the errors are relevant. However, we find that the two simulated GW data sets with two different fiducial $H_0$ values yield different results for the constraint uncertainties on the IDE models. This shows that the central values of parameters  can actually make some impacts on the constraint errors, and thus they are somewhat meaningful in this sense. Therefore, we still list the best-fit values from the simulated data in table \ref{results_table}, but it is important to keep in mind that only the constraint errors are meaningful for a forecast study.

Here, we denote the results from the simulated GW data with the fiducial $H_0=67.4~\mathrm{km~s^{-1}~Mpc^{-1}}$ as case 1 and denote the one with the fiducial $H_0=73.2~\mathrm{km~s^{-1}~Mpc^{-1}}$ as case 2. In both cases, we can see that the constraint on $H_0$ with GW data alone is tight and comparable with that from CMB+BAO+SN because GWs provide the measurements of absolute distances with high sensitivity to $H_0$. For $\Omega_{\rm{m}}$ and $\beta$, GW alone gives a weaker constraint than CMB+BAO+SN. However, the parameters degeneracy directions from GW and CMB+BAO+SN are rather different, especially in the $H_0-\Omega_{\rm{m}}$ and $H_0-\beta$ planes. Therefore, their combination could effectively break the degeneracies between parameters and give tight constraints on all parameters. With respect to the $\beta$ describing the strength of the interaction between dark energy and dark matter, we find that the data sets CMB+BAO+SN can give a tight constraint on it, $\beta=0.0003\pm{0.0036}$, which suggests that there is no interaction between dark energy and dark matter. Since the interaction term is proportional to $H(z)$, the CMB measurements in the early universe have a relatively strong ability to constrain the dimensionless coupling parameter $\beta$. Conversely, the GW measurements of the late universe have a weak constraint ability to it as shown in figure \ref{lcdm1}. However, GW combined with CMB+BAO+SN is still beneficial in breaking the degeneracy between $\beta$ and other parameters.

Comparing the two cases, we find that the constraints on $H_0$ are very different, whether using GW data alone or combined with CMB+BAO+SN, which indicates that GW data play a dominant role in constraining $H_0$. Moreover, since the degeneracy directions of GW and CMB+BAO+SN are quite different, the best-fit values of parameters from the combined CMB+BAO+SN with two GW cases, respectively, should be affected. For example, CMB+BAO+SN+GW (case 1) gives a result of $\Omega_{\rm{m}}=0.3163\pm0.0029$, while a shifted best-fit value of $\Omega_{\rm{m}}=0.2546\pm0.0023$ is obtained by the data set of CMB+BAO+SN+GW (case 2). In particular, the constraints on the interaction parameter $\beta$ are also greatly affected. The CMB+BAO+SN+GW (case 1) data give a result of $\beta=-0.0003\pm0.0032$ in excellent agreement with the zero interaction, while the result of $\beta=-0.0152\pm0.0022$ from the CMB+BAO+SN+GW (case 2) data supports that there is an interaction between dark energy and dark matter. Therefore, we conclude that an accurate measurement of $H_0$ is very helpful to explore the interaction between dark energy and dark matter.

For the I$w$CDM1 model, we find that CMB+BAO+SN can offer tight constraints on all parameters, while GW data only give weak constraints on all parameters except $H_0$. However, all parameters can be constrained more stringent with the help of combining with GW data. For instance, the constraint precision of $H_0$ and $\Omega_{\rm{m}}$ is improved by a factor of two when the CMB+BAO+SN data combine with GW. Concerning the coupling constant $\beta$, a tight constraint from existing CMB+BAO+SN, $\beta=0.0012\pm{0.0045}$, also indicates that there is no interaction between two dark sectors.  When combined with GW (case 1), we find that the constraint on the interaction parameter $\beta$ is not significantly improved, with a result of $\beta=0.0008^{+0.0045}_{-0.0041}$. However, when combined with GW (case 2), the error of $\beta$ improves to $0.0029$, but its best-fit value shifts to 0.0131, which means dark matter decay into dark energy.

The results of I$\Lambda$CDM2 model with $Q=3\beta H\rho_{\rm{c}}$ are shown in figure \ref{lcdm2} and table \ref{results_table}. Similar to the case of the I$\Lambda$CDM1 model, GW data alone give weak constraints on all parameters except $H_0$, while all parameters can be constrained very well by the CMB+BAO+SN data. The CMB+BAO+SN data combined with GW could significantly break the degeneracies between parameters and give tighter constraints on all parameters. For the coupling parameter $\beta$, we obtain $\beta=0.0003\pm{0.0015}$ from the CMB+BAO+SN and $\beta=0.0003\pm{0.0009}$ from CMB+BAO+SN+GW (case 1). With the addition of the GW data (no matter case 1 or case 2), the constraint precision of $\beta$ is improved by a factor of two. In this paper, we improve the simulation of GW to make it more reasonable and realistic. We note that our results are very close but slightly different from the previous results for the same IDE model. With respect to the coupling parameter $\beta$, Li et al. \citep{Li:2019ajo} obtained $\beta=0.00120\pm{0.00088}$ from the CMB+BAO+SN+GW data, in which the GW data are simulated with the fiducial $H_0=67.4~\mathrm{km~s^{-1}~Mpc^{-1}}$. It should be noted that the form of the interaction term they adopt is $Q=\beta H\rho_{\rm{c}}$, a factor of 3 different from the one we used. Therefore, the precision of their constraint on $\beta$ translated into our form should be 0.00029, a little better than that we obtained. The reason for this is that the number of GW standard sirens we obtained is only 400 by taking into account the coincident GRB detection, while Li et al. \citep{Li:2019ajo} just roughly assume 1000 GW standard sirens.  In addition, for case 2, the CMB+BAO+SN+GW (case 2) data give a result of $\beta=0.0095\pm0.0006$, which is deviated from zero.

Finally, we investigate the I$w$CDM2 model with $Q=3\beta H\rho_{\rm{c}}$, of which the results are shown in figure \ref{wcdm2} and table \ref{results_table}. As the number of parameters increases, we find that GW alone has a weak ability to constrain parameters. However, its parameter degeneracy direction is rather different from that of the CMB+BAO+SN data. Therefore, the combination of them could significantly improve the constraint of parameters. Compared with the constraint value of $\beta$ in the previous work \citep{Li:2019ajo}, $\beta=-0.0005\pm 0.0004$ from CMB+BAO+SN+GW, the error of $0.0009$ from  CMB+BAO+SN+GW (case 1) is also slightly large. Interestingly, in case 2, the combination of CMB, BAO, SN, and GW (case 2) yields a constraint on $\beta$ with a negative mean value, $\beta=-0.0089\pm0.0006$, which strongly supports dark energy would decay into dark matter.

\section{Conclusions}

Recently, the measurement inconsistencies of some key cosmological parameters imply that our understanding of the universe may be flawed under the framework of standard cosmological model, which urges us to reexamine some fundamental issues, e.g., the interaction between dark matter and dark energy. On the other hand, in a new era of multi-messenger astronomy, abundant standard sirens from the BNS mergers will be observed by the third-generation ground-based GW detectors, which will bring significant implications for cosmology. In this paper, by considering the coincidences between GWs and GRBs, we construct a mock catalogue of standard sirens consisting of about 400 events based on a 10-year observation of the ET and THESEUS missions. A more reasonable and realistic prospect for constraining the IDE models with future GW standard siren observation is presented in this paper.

We consider four IDE models, i.e., the I$\Lambda$CDM1 ($Q_1=3\beta H \rho_{\mathrm{de}}$), the I$w$CDM1 ($Q_1=3\beta H \rho_{\mathrm{de}}$), the I$\Lambda$CDM2 ($Q_{2}=3 \beta H \rho_{\mathrm{c}}$), and I$w$CDM2 ($Q_{2}=3 \beta H \rho_{\mathrm{c}}$). As we know, there is severe tension on measurements of $H_0$ between the early universe and the late universe. To discuss how robust the constraints on the IDE models from GWs are when the fiducial cosmology is altered, we simulate the GW data based on two different fiducial $H_0$ values, respectively, and discuss their constraints on IDE models. In addition to the simulated GW data, we use existing CMB, BAO, and the latest SN Ia Pantheon+ compilation. From obtained results in these four IDE models, in summary, GW data play a similar role. GW data alone cannot constrain tightly on all parameters except $H_0$. However, it can be an important supplement to the CMB+BAO+SN data. The parameter degeneracy directions from GW and CMB+BAO+SN are rather different. Therefore, the combination of them could effectively break the degeneracies between parameters and give much tight constraints on all parameters. For the dimensionless coupling parameter $\beta$ describing the strength of the interaction between dark energy and dark matter, we find that the constraint results from the existing CMB+BAO+SN data indicate no interaction between two dark sectors. In the form of $Q_1=3\beta H \rho_{\mathrm{de}}$, we find that the addition of GW standard siren data improves little for the constraint on $\beta$. While, in the form of $Q_{2}=3 \beta H \rho_{\mathrm{c}}$, the constraint precision of $\beta$ is improved by a factor of two with the addition of the GW data. We conclude that the GW data could play a more important role in determining the interaction between dark energy and dark matter in the models with $Q=3 \beta H \rho_{\mathrm{c}}$, compared with the models with $Q=3\beta H \rho_{\mathrm{de}}$. More interestingly, comparing two constraints on $\beta$ from two simulated GW data sets based on $H_0=67.4~\mathrm{km~s^{-1}~Mpc^{-1}}$ and  $H_0=73.2~\mathrm{km~s^{-1}~Mpc^{-1}}$ respectively, we find that the CMB+BAO+SN+GW (case 2) data yield a non-zero value in all four IDE models indicating the presence of an interaction between dark energy and dark matter.  Therefore, we conclude that an accurate measurement of $H_0$ is helpful to explore the interaction between dark energy and dark matter.

In summary, there is no doubt that the detection of GW will greatly promote the development of modern cosmology, and GW data are expected to be significantly helpful in answering fundamental issues, such as the interaction between dark energy and dark matter. We also pin hopes on the second-generation space-based GW detector, such as DECi-hertz Interferometer Gravitational-wave Observatory (DECIGO), sensitive to the frequency range between target frequencies of the Laser
Interferometric Space Antenna and ground-based detectors \citep{2022ApJ...931..119Z,2020ApJ...905...54G,2021EPJC...81...14Z,2021ApJ...908..196P,2022ApJ...926..214C}. In the time frame considered in this work, numerous EM surveys are planned for construction and operation. For instance, the next-generation CMB experiments, Simons Observatory \citep{SimonsObservatory:2018koc} and CMB-S4 \citep{CMB-S4:2016ple}, will accurately measure the temperature and polarization anisotropies of CMB at multiple wavelengths with unprecedented accuracy. The Dark Energy Spectroscopic Instrument (DESI) \citep{DESI:2016fyo} and the Square Kilometre Array (SKA) \citep{Maartens:2015mra,Zhang:2021yof} will precisely measure the BAO and the large-scale structure of the universe at optical and radio frequencies, respectively. The Vera Rubin Observatory Legacy Survey of Space and Time (LSST) is expected to process $\sim 10^6$ transient detections per night, increasing the SNIa sample size by up to a factor of 100 compared to previous samples \citep{LSSTDarkEnergyScience:2021laz}. These future surveys will undoubtedly provide even more stringent estimations of the cosmological parameters, which makes us optimistic that we can finally confirm whether there is an interaction between dark energy and dark matter.

\acknowledgments
This work was supported by the National SKA Program of China (Grant No. 2022SKA0110200 and 2022SKA0110203) and the National Natural Science Foundation of China (Grants Nos. 12205039, 11975072, 11835009, and 11875102).

\bibliography{IDE_ref}
\bibliographystyle{JHEP}

\end{document}